\documentclass{article}

\usepackage[english]{babel}

\usepackage[letterpaper,top=2cm,bottom=2cm,left=3cm,right=3cm,marginparwidth=1.75cm]{geometry}

\usepackage{amsmath}
\usepackage{graphicx}
\usepackage{hyperref}
\hypersetup{colorlinks,allcolors=black}
\title{Game Dynamics Structure Controlled by Design: \\
an Example from Experimental Economics}
\author{Wang Zhijian \footnote{The author thanks Wang Yijia and Pan Gang for the critical idea, comments and lasting supporting during this work carry out. The author thanks Zhang Jianbo for his patience on the proofreading. All the error belongs to the author.} \\ Experimental social science laboratory, Zhejiang University, China}

\begin{document}
\maketitle

\begin{abstract}
Game dynamics structure (e.g., endogenous cycle motion) 
in human subjects game experiments can be predicted by game dynamics theory. 
However, whether the structure can be controlled 
by mechanism design to a desired goal is not known. 
Here, using the pole assignment approach in modern control theory,
we demonstrate how to control the structure 
in two steps: 
(1) Illustrate an theoretical workflow on 
how to design a state-depended feedback controller for desired structure;  
(2) Evaluate the controller by laboratory human subject game experiments and by agent-based evolutionary dynamics simulation. 
To our knowledge, this is the 
first realisation of the control of the human social game dynamics structure in theory and experiment. 
\end{abstract}

\tableofcontents

\section{Introduction} 

\subsection{Background}
\begin{figure}[h!]
\centering
\includegraphics[width=0.9\textwidth]{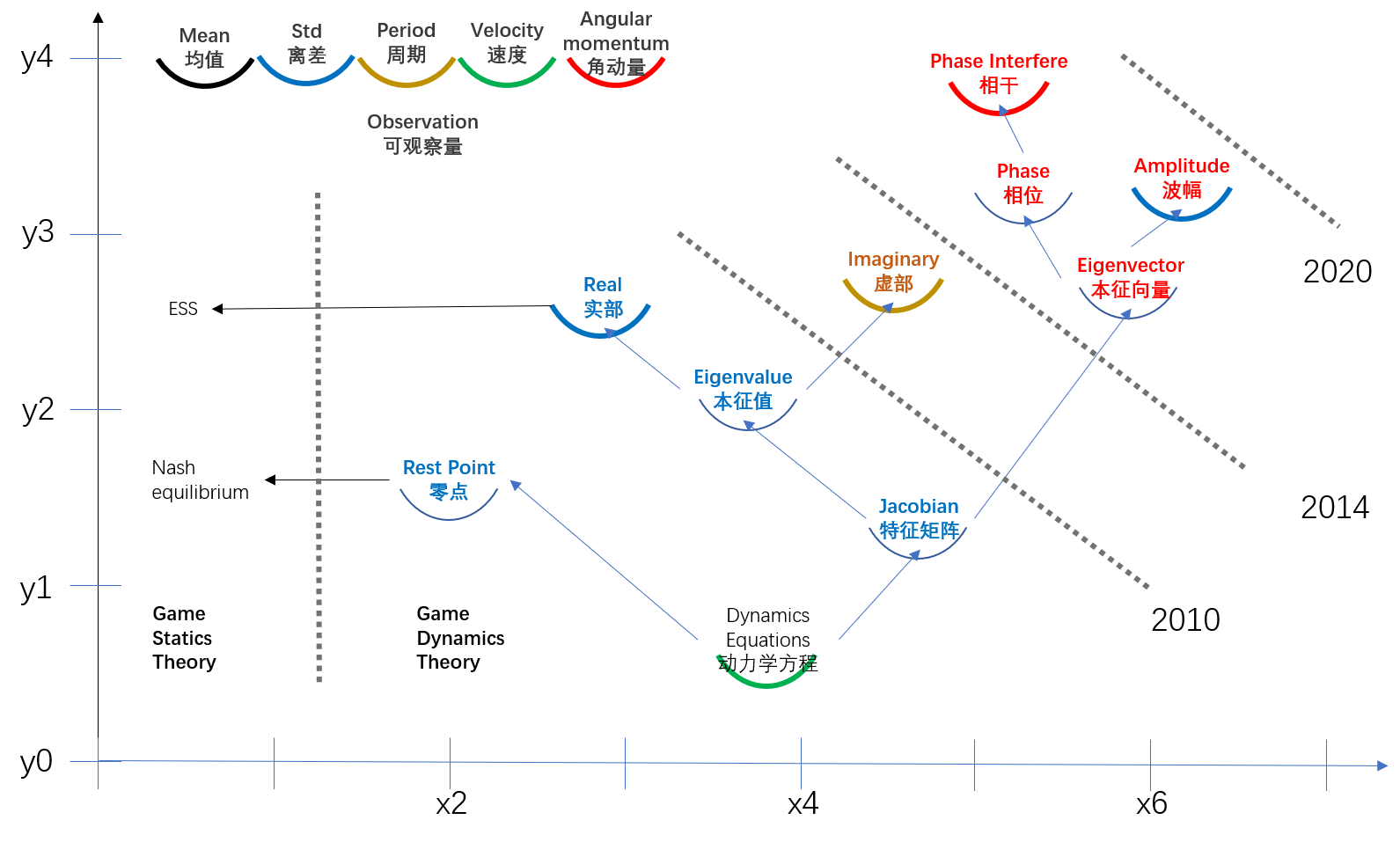}
\caption{Knowledge tree of game dynamics theory. This knowledge tree is referring to the basic mathematics concept.  This map guides the development of game dynamics theory and experimental observations in the past decade (during 2010-2021). \label{fig:tree2}} 
\end{figure}

Theory and experiment are the two sides of a coin of science. 
Game theory is a science disciplines about strategy interactions. 
Borrowing from basic dynamics system theory in mathematics, we can use a figure to illustrate the background of current  situation of
game dynamics theory, see Figure \ref{fig:tree2}.

The static version of game theory (call also as classical game theory or game statics theory) has well developed in past seven decades.  It has clear central concept (Nash equilibrium), clear examples in experiments (behaviour game experiments \cite{Behavioral2003}) and in application of policy mechanism design \cite{wiki:Mechanism_design}. The static version provides the outcome of a game, but not how the agents play a game. The picture of this version is noteworthy in that equilibrium concept come into being divorced from a dynamic process \cite{samuelson2016,pangallo2019best}. 

The dynamics version of game theory is called as evolutionary game theory or as game dynamics theory \cite{smith1982evolution,2011Sandholm}. It puts the dynamic process back into the full picture of game theory. This version initially surrounded by a great deal of excitement, because it is a novel paradigm to describe the evolution in widely various field, from social science to nature science. For a while (approximately form 1980s to 1990s), it lay at the center of game theory as well as perhaps economic theory more generally \cite{samuelson2016}. 
For long time, the dynamics version remained in  qualitative argument or used of very fundamental topological facts, like (1)
 Rest point, shown at [$x$:2, $y$:1.5] in Fig. \ref{fig:tree2}, which links to the Nash equilibrium (shown at [$x$:0.5, $y$:1.5]) concept.   
(2) ESS (evolutionary stable strategy) concept, shown at [$x$:0.5, $y$:3.5] in Fig. \ref{fig:tree2}, which actually also links to the Nash equilibrium (shown at [$x$:0.5, $y$:1.5]) concept. It plays as necessary condition to distinguish whether a rest point can be a strict Nash equilibrium. The real part of the eigenvalues [$x$:3, $y$:2.5] are concerned (e.g., \cite{2021continuous}) as qualitative iteration for the stability.     
These  concepts lack of clear exemplified evident with accurate quantitative verification. 
For game dynamics theory, its central concept set  has not well established, and its application (policy design) is rarely seen, 
and the accurately merging of the theory and real experiments observation is rarely seen either. So, for long time, game dynamics theory plays as an assistance role for game static theory.

The past decade has seen obviously improvement. The experimental observations of the dynamics structure (e.g., the endogenously cycles in Rock-Paper-Scissors and in matching pennies game experiments near 2014 \cite{dan2014,wang2014social,wang2014}),  which has been long expected since the established the replicator dynamics \cite{taylor1978evolutionary} for evolutionary game theory, was test out. Further, in various experimental setting, not only the stability, but also the quantitative relationship like frequencies of cycles (reflect the imaginary part of the eigenvalue of the dynamics, at [$x$:4.5, $y$:2.5] near 2014 \cite{wang2014}), the velocity field (near 2010 \cite{2011coyness}) and angular momentum \cite{wang2017} 
were test out. Further more, in high dimensional games, the invariant manifold (reflect the inner component structure of the eigenvector of the dynamics) become clear in experiments \cite{WY2020,2021Shujie,2021Qinmei} (in fig. \ref{fig:tree2}, at [$x$:6, $y$:3]  near 2020). These results indicate that, like the equilibrium concept in the static version of game theory, the dynamics  structure (or called as dynamics pattern, or called as invariant manifold) appears as the focus concept for the dynamics version of game theory. 

\subsection{Research question}
Having the obviously improvement of the past decade, naturally, an engineering question rise: Whether the dynamical structure can be controlled?  This is the research question of this study. To the best of our knowledge, it has never been realised before. We hope to provides a simple and clear example to show how to control dynamics structure by two steps: (1) design and (2) verify by the laboratory human game experiments.  

On the control-by-design of the dynamics structure, 
our idea mainly based on following two aspects: 
\begin{itemize}
    \item \textbf{Game dynamics structure can be predicted by the linear dynamics system approximation, where the eigenvector plays a cruel role. } --- Game experiment data support the suggestion that, the dynamics structure is governed by the complex eigenvector components in the dynamics equations (e.g., replicator dynamics) and its linearization near the (Nash) equilibrium. Previous results has shown that, even in discrete time and discrete strategy game, the linearization approximation works. Examples come from 4-strategy one-population games (Zhou 2021 \cite{2021Shujie}),  5-strategy one-population games (Yao 2021 \cite{2021Qinmei}), and 4 strategy two-role zero-sum asymmetry game of the O'Neill 1987 game (\cite{ONeill1987}, Wang and Yao 2020 \cite{WY2020}). These results show that, by the eigencycles (or invariant manifold, basing on the eigenvectors which being out of the linear approximation) measurements,  the game dynamics theory can capture human subject experimental dynamics structure exactly.  This logic chain can be seen in Figure \ref{fig:tree2} as  chain from Jacobian to eigenvector and eigencycles (phase interfere).   
    \item \textbf{A linear dynamics system is controllable} --- Apply mathematics shows that, invariant manifold concept provides a clear picture for dynamics process description.
    This concept roots in dynamical systems theory \cite{wiki:Dynamical_systems_theory}, a solid branch of mathematics. Dynamical systems deal with the study of the solutions to the equations of motion of systems which are primarily mechanical in nature. For dynamics system of first-order ordinary differential equations system, the state-space method for feedback controller design is a well studied and applied engineering field, namely as modern control theory and application \cite{2015feedback,2010moderncontrol}. 
    Considering the state depend feedback consequences on game dynamics process 
    has been noticed (e.g.\cite{2015evolutionary,2020FuFeng}), 
    although the feedback is not of control-by-design in existed literature.  
\end{itemize}
 For this study, these two aspects are the background of the main logic chain. 

\subsection{Main logic}
This study illustrate how to control the structure by a clear example of one symmetry 5 strategy game. We control an eigenvalue to change the eigenvector, then to change the dynamics structure. As shown in Figure \ref{fig:tree2}, control the real part of an eigenvalue  (at $x$:3,$y$:2.5) to desired aim, to investigate the change  eigenvector components (at $x$:6,$y$:2.5) and its observable  (the eigencycles at $x$:5.5,$y$:4) should also be changed. When controlling the eigenvalue, we use the pole assignment for linear system approach at the Jacobian (at $x$:6,$y$:2.5). This is the outline of the logic. 

The main technical point of this work is the workflow for the controller design, which is introduced in the next section.  
In Section 3, we will practically show an example to realise the workflow suggested above. 
Then, by agent-base simulations and by  laboratory human behaviour game experiment, we will verified the theoretical prediction of the designed controller. 
In Section 4, we summarise the results and point out the related concept and further directions.

\section{Controller design by pole assignment}
\subsection{Workflow}
     In this paper, our control-by-design (mechanism design) approach comes from \textbf{single-input pole assignment} approach  for linear system in the modern control theory. It can be realised in following workflow (we will show how to realise the control by 5-strategy  game as an example in the next section):
    \begin{enumerate}
    \item Solve the dynamics equation for the original Jacobian ${J^o}$ and the eigen-system (eigen-value and eigen-vector ${x_o}$);
    \item Assign controllable channel vector ${B}$ and the desired eigenvalue; Meanwhile, the necessary condition, as well as the optimal goal of the controller needs to be clarified.
    \item Construct the controlled Jacobian ${J^c}$ from the original Jacobian ${J^o}$, the control vector ${K}$ and the channel vector ${B}$;
    \item Solve the ${K}$ to make the eigenvalue of the ${J^c}$ being identical (or the closest) to desired eigenvalue. 
    \item Provide clear theoretical prediction of the controller consequence. Notice that, the designer needs to clarify the measurement to test the consequence. The consequences includes whether the goal of design has reached, whether the constrain condition is satisfied, and whether the system make sense (appears reasonable) when controller rule is applied. 
    In the pole assignment of single-input linear systems using state feedback technology, as having the solved ${K}$ for given $B$, one can evaluate the consequence of the control. 
    \item Verify the reality of step-5 theoretical predictions by simulation and experiment 
        \begin{enumerate}
            \item numerical (agent-based) computer simulation
            \item laboratory human subject game experiment
        \end{enumerate} 
\end{enumerate}

\subsection{Algorithm for the controller design}
~\\
\textbf{[Basic aspect]} Assume that, an one population $n$ strategy   game has unique mixed strategy Nash equilibrium, whose dynamics can be described by the replicator dynamics, 
\begin{equation}\label{eq:rp}  
\dot{x_i} = x_i \cdot \big(U_i - \Bar{U}\big). 
\end{equation}
This is a time-invariant systems, because time $t$  does appear as a variable in the rhs terms. Assume again that, the game system can be expanded as 
\begin{equation}\label{eq:jo}  
\dot{x}={J^o} \cdot x,
\end{equation}
wherein, ${J^o}$ is the Jacobian matrix evaluated at Nash equilibrium $x_{\text{Nash}}$, and $x$ (a $1 \times n$ vector, a row vector) is a small deviation from $x_{\text{Nash}}$. 

In mathematics, the dynamics is specified on the $n$-simplex by an ordinary
differential equation (ODE). In control engineering, this is a continuous-time LTI   (linear time-invariant) system, which is a fairly common situation to carry out feedback control (see wiki: State-space\_representation). 
~\\
~\\
\textbf{[Construct ${J^c}$]} We design the control term following linear control approach, by adding  $K$ control vector (an $1 \times n$ vector, a row vector)  on the channel $B$ (a $n \times 1$ vector, a column vector), and in addition, a financial balance term $T$. Then the controlled system can be expressed as 
    \begin{equation}\label{eq:BKT} 
          \dot{x}={J^o} \cdot x + {B} \cdot {K} \cdot x + T \cdot x
    \end{equation}
Here, $B \cdot K$ is the ordinary linear state dependent control term (see standard text book of modern control system). The financial balance term $T := - \sum_{i=1}^n(B \cdot {K} \cdot x) $ is a constant at the given state. Then, the controlled Jacobian ${J^c}$ can be expressed as
\begin{equation}\label{eq:jc1}   
{J^c} = {J^o} + {B} \cdot {K} + T.
\end{equation}
Then for an assigned channel $B$, once ${K}$ is given , ${J^c}$, as well as the eigenvalues and associated eigenvectors ${x_c}$, are determined. Then, the 4-th step above is realisable, and the desired dynamics structure | the associated eigencycles ${\sigma_c}$ | can be expected. 
~\\
~\\
\textbf{[Explain terms in ${J^c}$]} 
 For the simplicity in this study, we assume that, the control is a single channel (single-input) control system. That is, the assigned channel vector ${B}$ is an unit vector, 
like [0,...1,...,0]$^T$, and the non-zero element appears in the $m$-dimension indicating the control is assigned at the single $m \in \{1,2,...,n\}$ channel. The pole assignment problem for a single-input controllable system is relatively straightforward to solve. 
So, we illustrate by solving the pole assignment problem for single-input systems. 
Then, we have following two properties on the construction:
\begin{enumerate}
\item When $K \cdot x_{\text{Nash}} = 0$  is satisfied, the equilibrium of the system remained at  the Nash equilibrium $x_{\text{Nash}}$ of original game, and is preserved. 
\item Referring to the definition of $T$, the control do not influence the total payoff of the game. That is to say, the financial conversation is preserved.
The payoff redistribution according to following two parts
\begin{enumerate}
\item Reward the strategy on the Channel $j$: Total reward value is ${BKx}_j$. 
Then, for each individual agent on this channel, its reward (added payoff) is ${BKx} / x_j$.  
\item Tax each strategy channel: Tax value is $T$. Then, for each individual agent on its channel $k \in \{1,2,...,n\}$, its tax value is $T$. Notice that, the reward channel strategy is not excluded to be taxed too. 
\end{enumerate}
\end{enumerate}
\textbf{[On Constraint condition of the controller design]}  
\begin{enumerate} 
\item The equilibrium state in the strategy space should be preserved and not be shifted; This can be satisfied by setting $K \cdot x_{\text{Nash}} = 0$; and naturally  the equilibrium payoff will not be shift;
\item There is no additional payoff (financial support) 
adding to the game system, which naturally can be realised by the definition of $T$.   
\end{enumerate} 

\section{The Y5 game as example}

We employ a five strategy symmetry game to show how to observe the eigensystem and how to control the dynamics structure in experiment. 
The game is an one population 5$\times$5 symmetric game. In strategy state space, its evolution trajectory is of a 5 dimensional trajectory. Table \ref{tab:gamemodel}
shows the 5-strategy payoff matrix:
\begin{table}[h!]
\caption{The Y5 game matrix}
\begin{center}
\begin{tabular}{c|rrrrr}
 \hline 
&x1&x2&x3&x4&x5\\
 \hline 
x1&0&3&4&11&11\\
x2&5&0&2&11&12\\
x3&2&5&0&9&12\\
x4&6&10&10&0&3\\
x5&10&10&10&4&0\\
 \hline
\end{tabular}
\end{center}
\label{tab:gamemodel}
\end{table}

 This game matrix is borrowed from  Yao \cite{2021Qinmei} rooting in Freidlin and Wentzell (Chapter 6.6)
 \cite{fre1994} and Newton \cite{Newton2018}. The underlying idea is simple. When  assume that the values in matrix as cost, the dynamics without perturbations converges to the two set of the strategies \{$x_1,x_2,x_3$\} and \{$x_4,x_5$\}, as shown in \cite{fre1994}\cite{Newton2018}. Alternatively, when assume that the values in matrix as reward, the dynamics without perturbations could converges to the unique mixed Nash equilibrium, because all of the real part of the eigen values are negative. There has a unique pair of complex eigenvalues; And its dynamical structure is determined by the eigenvector associated to the unique pair of complex eigenvalues, which has been shown
 by Yao \cite{2021Qinmei}. 
 
 We use this game as a benchmark.  
Because Yao \cite{2021Qinmei} has shown us that: The dynamics structure, game dynamics theory and data meet well. The 5 dimensional dynamics structure in human game experiments and in agent-based simulation with reinforcement learning model can be predicted by the linear approximation of the evolutionary game dynamics theory. In this study, we use also the data (human game experiments $E^o$, simulation $S^o$  and theoretical $T^o$ in Table \ref{tab:numResult}) from  Yao \cite{2021Qinmei}  as reference to illustrate how to control the dynamics structure.  

\subsection{Step 1: Evaluate the original eigen system}
To investigate dynamics behavior in a laboratory experiment game,
we use the replicator dynamics equation \cite{2011Sandholm}:
\begin{equation}\label{eq:repliequl} \Dot{x}_j=x_j\Big(U_j - {\overline{U}}\Big),
\end{equation}
$x_j$ is the $j$th strategy player's probability
in the population with the $j$th strategy player included, and
$\Dot{x}_j$ is the evolution velocity of the probability;
 $U_j$ the payoff of the $j$th strategy player;
And $\overline{U}_X$ is the average payoff of
the all population, which equals to $\sum_{k=1}^{5} x_k U_k$. 
This 5-dimension space has one concentration 
$ \sum_j x_i=1 \cap x_i \geq 0 ~ (i \in \{1,2,...,5\})$.
An explicitly expression of these nonlinear differential equations 
are shown in SI in Yao \cite{2021Qinmei}.  
Then, at any time, the system must be
\textcolor{red}{a} five-dimension space simplex. 
It is easy to verified that, the unique Nash equilibrium is
\begin{equation}\label{eq:Nash_eq}
x^*_{\text{Nash}} =(x_1^*,x_2^*,x_3^*,x_4^*,x_5^*) = \frac{1}{2987}(444, 638, 641, 288, 976). 
\end{equation} %
 and the value of the game is $18382/2987 = 6.1540$, which is the payoff of each strategy at equilibrium. 
It can also be verified that, $x^*$
is the unique rest point of which the velocity is zero and none of the strategy density be 0, and of which is identical to the unique Nash equilibrium $x_{\text{Nash}}$ of the game. Then,  the Jacobian matrix ${J^o}$ evaluated at Nash equilibrium $x^*_{\text{Nash}}$ is

$$ {J^o} =  \left[
     	\begin{array}{ccccc}
     	 -1.7090 & -1.3236 & -1.1011 & -0.3533 & -0.3295 \\
     	 -1.3877 & -2.5427 & -2.0094 & -0.5077 & -0.2599 \\
     	 -2.0381 & -1.4817 & -2.4480 & -0.9393 & -0.2611 \\
     	 -0.5300 & -0.1836 & -0.1357 & -1.2898 & -0.9851 \\
     	 -0.4892 & -0.6223 & -0.4599 & -3.0639 & -4.3185 \\
     	\end{array}
     \right] $$
And its eigenvalues $\lambda^o$ and eigenvectors $v^oo$ are
\begin{equation}\label{eq:lambda_o}
    \lambda^o=\left[
     	\begin{array}{ccccc}
     	 -6.154 & 0 & 0 & 0 & 0 \\
     	 0 & -4.351 & 0 & 0 & 0 \\
     	 0 & 0 & -0.667 + 0.429i& 0 & 0 \\
     	 0 & 0 & 0 & -0.667 - 0.429i& 0 \\
     	 0 & 0 & 0 & 0 & -0.468 \\
     	\end{array}
     \right]
\end{equation} 
     
     
    $$v^o=\left[ 
     	\begin{array}{rrrrr}
    ~~0.310  &  -0.224  &     0.424\exp(         0.000\pi i)  &   0.315 - 0.284i &   0.061 \\
    0.446  &  -0.399  &   0.525\exp(   -0.464\pi i)  &   0.394 + 0.346i &  -0.402 \\
    0.448  &  -0.353  &  0.636\exp(    0.766\pi i)   &  -0.636 - 0.000i &   0.465 \\
    0.201  &   0.178  & 0.301\exp(    0.383\pi i)  &  -0.108 - 0.281i &  -0.615 \\
    0.682  &   0.797  &  0.222\exp(   -0.683\pi i)   &   0.036 + 0.219i &   0.491 \\
     	\end{array}
     \right]$$ 
It is not surprise that, the most negative eigenvalue is the minus of the value of the game. This eigenvalue relates to the first column eigenvector which represented 
equilibrium distribution (see p120 in \cite{dan2016}). As all the real part of the eigenvalue is negative, the game is asymptotically stable. 
     
The 3rd column of the eigenvector matrix is the original complex eigenvector which  
 represents the dynamics structure.  
In this example, the third eigenvalue is the object of control-by-design, and its related eigenvector is the object of the observation (the eigencycle).

\subsection{Step 2: Assign control goal}\label{sec:assigngoal}

In this study,  to illustrate controllability of the game dynamics structure, we simply assign to a single channel.  
That is, the assigned channel vector ${B}$ is an unit vector. We suppose that,  
$$ {B} = \big[0  ~~0~~0~~0~~1\big]^T.$$ 
The necessary and sufficient condition for the controllability of a time invariant linear dynamics system is that the matrix 
$$\begin{bmatrix} {B} & {J^o} {B} & {J}^{2}_{o} {B} & \cdots & {J}^{n-1}_{o} {B}\end{bmatrix} $$  
is non singular and is full rank (that is  $n$ = 5). It can be verified that, this matrix is full rank, 
and the original game is controllable. Why we choose this channel  will be explained  in appendix (see Section \ref{app:designgoal}). 
~\\
~\\
The design is based on pole assignment using state feedback with the desired closed-loop poles at the complex eigenvalues $\lambda^o(3)$:=$-0.6674 + 0.4294i$ and its complex conjunction $\lambda^o(4)$=$-0.6674 - 0.4294i$. Our goal is to reduce or  increase its real part of the eigenvalue.  To present it completely,
\begin{equation}\label{eq:design_goalm}
    ~{\lambda^-} =  ~{\lambda^o}(3) - b =  ~{\lambda^o}(3) - 0.2 = -0.8674 + 0.4294i
\end{equation} 
\begin{equation}\label{eq:design_goalp}
    ~{\lambda^+} =  ~{\lambda^o}(3) + b =  ~{\lambda^o}(3) + 0.2 = -0.4674 + 0.4294i
\end{equation} 
Here, $b:=0.2>0$ is a real value (Why we choose $b=0.2$ will be explained in appendix \ref{app:designgoal} in details), the superscript $o$ indicates the original Y5 game 
which is the benchmark game. The superscript $-$ (or $+$) represent the  controlled treatment which the desired real part of the eigenvalue is minus (or plus).  
  
The experimental observable consequence 
of the desired goal will be shown in  Section \ref{subsec:theoPred}. Referring to the design goal shown in appendix \ref{app:designgoal}, the criterion is that ---
The dynamics structures of the two controlled treatments differ, but both can meet the their theoretical predictions respectively; And at the same time, the  dynamical structure of the two controlled treatments are irrelevant although they rooted in the same original Y5 game.   
 
~\\

\subsection{Step 3: Construct the controlled Jacobian ${J^c}$}

Referring to the Eq. \ref{eq:BKT}, we can reach that, 
 \begin{equation}\label{eq:jk1} \left[\begin{array}{c} \dot{x}_1 \\ \dot{x}_2 \\ \dot{x}_3 \\ \dot{x}_4 \\ \dot{x}_5 \end{array}\right] =
{J^c} \left[\begin{array}{c} x_1 \\ x_2 \\ x_3 \\ x_4 \\ x_5 \end{array}\right] = {J^o} \left[\begin{array}{c} x_1 \\ x_2 \\ x_3 \\ x_4 \\ x_5 \end{array}\right] + 
\sum_{i=1}^{5}{k_i x_i} \left[\begin{array}{c} 0 \\ 0 \\ 0 \\ 0 \\ 1\end{array}\right] 
-  \sum_{i=1}^{5}{k_i x_i} \left[\begin{array}{c} x_1 \\ x_2 \\ x_3 \\ x_4 \\ x_5 \end{array}\right] 
 \end{equation}
Notice that, as the relation --- $\sum_{i=1}^{5}{k_i x_i^*} = 0$  --- must hold. It is important, because the controller is required to preserve the  equilibrium.   That is to say, of the two controlled treatments, their strategy distributions  remain the same as that of the original Y5 game shown in Eq. \ref{eq:Nash_eq}.   
Having the ${J^c}$, we can solve pole assignment problem for linear system.

\begin{figure}
\centering
\includegraphics[width=1\textwidth]{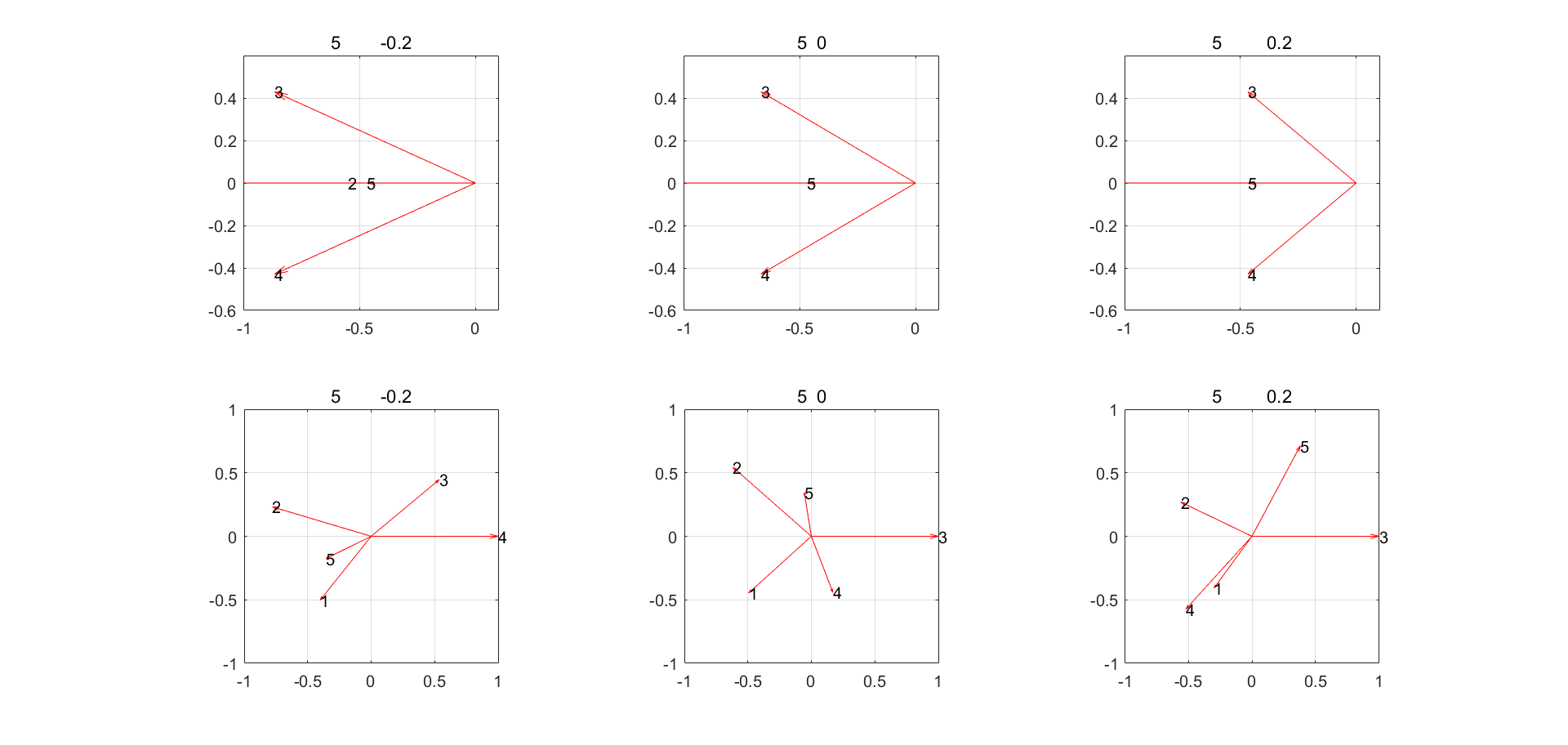}
\caption{\label{fig:eVector} The eigenvalues (the top row) and their associated three eigenvectors (the bottom row) represented in complex space, where the real part is in horizon and imaginary part in vertical. From left to right, the sub-figure comes from the $[-/o/+]$ treatment's Jacobian $[J^+/J^o/J^-]$ respectively. The number indicates the components by order along $(x_1,x_2,...,x_5)$. The complex eigenvalues and their related eigenvectors always come out as conjugate pairs but only one of them are really observable. Referring to \cite{2021Qinmei}, the eigenvector set related to the third eigenvalue is the observable in practice. One can verified by the phase angle of the eigenvector components of the subspace $\{x_1,x_2,x_3\}$ by seeing for example, the best response of $x_1$ is $x_3$ and of $x_3$ is $x_2$.}
\end{figure}

\subsection{Step 4: Solve the ${K}$ for desired goal}

In this study case,  the design goal is to control the dynamics structure. As mentioned above, The design goal can be expected to realised by the two treatments shown in Eq. (\ref{eq:design_goalm}) and Eq. (\ref{eq:design_goalm}). 

\begin{description}
  \item[1. For $b=-0.2$ treatment] which denoted as $[-]$-treatment, referring to Table \ref{fig:designgoal} in appendix \ref{app:designgoal}, we find that the numerical solution is 
  \begin{equation}
    {K_c^-}  =  \left(
     	\begin{array}{ccccc} 
     	k_1^- & k_2^- & k_3^- & k_4^- & k_5^-  \\
     	\end{array}
     \right) = \left(
     	\begin{array}{ccccc}
     	1.729 & 	-1.611 & 	0.454 & 	1.250 & 	-0.400   \\
     	\end{array}
     \right)    
  \end{equation} 
    $${J^-}=\left[
     	\begin{array}{rrrrr}
     	 -1.966 & -1.084 & -1.169 & -0.539 & -0.270 \\
     	 -1.757 & -2.199 & -2.106 & -0.775 & -0.174 \\
     	 -2.409 & -1.136 & -2.545 & -1.207 & -0.175 \\
     	 -0.697 & -0.028 & -0.179 & -1.410 & -0.946 \\
     	 0.675 & -1.707 & -0.154 & -2.223 & -4.588 \\
     	\end{array}
     \right]$$

And its eigenvalues $\lambda^-$ and their associated eigenvectors $v^-$ are
    $$\lambda^-=\left[
     	\begin{array}{ccccc}
     	 -6.154 & 0 & 0 & 0 & 0 \\
     	 0 & -4.351 & 0 & 0 & 0 \\
     	 0 & 0 &  -0.867 + 0.429i & 0 & 0 \\
     	 0 & 0 & 0 & -0.867 - 0.429i& 0 \\
     	 0 & 0 & 0 & 0 & -0.468 \\
     	\end{array}
     \right]$$
     

    $$v^-=\left[ 
     	\begin{array}{rrrrr} 
   0.310 & -0.224 &  0.391\exp(0.000\pi i) & -0.244 + 0.305i &  0.0612 \\
   0.445 & -0.399 &  0.493\exp(-0.377\pi i)  & -0.472 - 0.140i & -0.402 \\
   0.447 & -0.353 &  0.423\exp(0.936\pi i)  &  0.325 - 0.271i &  0.465 \\
   0.201 &  0.178 &  0.607\exp(0.715\pi i)  &  0.607  + 0.000i& -0.615 \\
   0.682 &  0.797 &  0.240\exp(-0.139\pi i) & -0.216 + 0.106i &  0.491 \\
     	\end{array}
     \right]$$

The most negative eigenvalue is the minus 
of the value of the game which is 6.1540, 
which remain identical to the original game. 
This eigenvalue relates to the first column eigenvector which represented the 
Nash equilibrium distribution, referring to Eq. \ref{eq:Nash_eq}. As all the real part of the eigenvalue is negative, the game is asymptotically stable. 
 
  \item[2. For $b$=+0.2 treatment] which denoted as $[+]$-treatment, referring to Table \ref{fig:designgoal} in appendix \ref{app:designgoal}, we can find that the solution is  
  \begin{equation}
    {K_c^+}  =  \left(
     	\begin{array}{ccccc} 
     	k_1^+ & k_2^+ & k_3^+ & k_4^+ & k_5^+  \\
     	\end{array}
     \right) = \left(
     	\begin{array}{ccccc}
     	-0.567 	& 1.362 	& -0.963 	& -1.357 	& 0.400   \\
     	\end{array}
     \right)    
  \end{equation} 
     
    $${J^+}=\left[
     	\begin{array}{rrrrr}
     	 -1.625 & -1.526 & -0.958 & -0.152 & -0.389 \\
     	 -1.267 & -2.834 & -1.804 & -0.218 & -0.345 \\
     	 -1.916 & -1.774 & -2.241 & -0.648 & -0.347 \\
     	 -0.475 & -0.315 & -0.043 & -1.159 & -1.024 \\
     	 -0.871 & 0.295 & -1.108 & -3.977 & -4.049 \\
     	\end{array}
     \right]$$
      
And its eigenvalues $\lambda^+$ and their associated eigenvectors $v^+$ are
    $$\lambda^+=\left[
     	\begin{array}{ccccc}
     	 -6.154 & 0 & 0 & 0 & 0 \\
     	 0 & -4.351 & 0 & 0 & 0 \\
     	 0 & 0 & -0.467 + 0.429i& 0 & 0 \\
     	 0 & 0 & 0 & -0.467 - 0.429i& 0 \\
     	 0 & 0 & 0 & 0 & -0.468 \\
     	\end{array}
     \right]$$
     
     
    $$v^+=\left[ 
     	\begin{array}{rrrrr} 
0.310  &  0.224  &   0.297\exp(0.000\pi i)    & -0.176 + 0.240i  & 0.061 \\
0.446  &  0.399  &   0.366\exp(-0.441\pi i)  & -0.330 - 0.158i  & -0.402 \\
0.448  &  0.353  &   0.588\exp(0.702\pi i)   &  0.588 + 0.000i  &  0.465 \\
0.201  & -0.178  &   0.455\exp(-0.033\pi i)  & -0.306 + 0.337i  & -0.615 \\
0.682  & -0.797  &   0.475\exp(-0.955\pi i)  &  0.224 - 0.419i  &  0.491  \\
     	\end{array}
     \right]$$  
The most negative eigenvalue is the minus of the value of the game, 
which remain identical to the original game. 
This eigenvalue relates to the first column eigenvector which represented 
equilibrium distribution (\cite{dan2016} p120), 
and the distribution of the controlled treatment 
is the same as the original game. 
As all the real part of the eigenvalue is negative, 
the controlled treatment is asymptotically stable.   
\end{description}

\subsection{Step 5: Define verifiable theoretical predictions of the controller}\label{subsec:theoPred}
Control the eigenvector would change the amplitude of vibration and the phase difference between various components of the eigenvector. These will lead to observable consequence referring to the original game. Due to the design aim, some of observation is invariant, and some changed.
The verifiable theoretical predictions of the controller are listed as following. 

\begin{description}
\item[1: Detect the distribution associates to the leading real eigenvalue]  The the strategy distribution of the controlled game should be identical to that of the original game, 
\begin{eqnarray}
 \rho^-:&=& (\rho_{1}^-, \rho_{2}^-, \rho_{3}^-, \rho_{4}^-, \rho_{5}^-) \\
   \rho^o:&=& (\rho_{1}^o, \rho_{2}^o, \rho_{3}^o, \rho_{4}^o, \rho_{5}^o) \\
   \rho^+:&=& (\rho_{1}^+, \rho_{2}^+, \rho_{3}^+, \rho_{4}^+, \rho_{5}^+) 
\end{eqnarray} 
Wherein the superscript indicates the $[-/o/+]$ treatment, the subscript indicates 
the (1,2,3,4,5)-strategy of the game respectively. 

On measurement, this prediction can be verified by the strategy proposition measurement in the time series. 
  \begin{eqnarray}\label{eq:meanrho}
\bar{\rho}_{i} =  \frac{1}{NT}\sum_{t=0}^T  c_{i}(t)  
\end{eqnarray}  
Herein, $N$ is the total number of agents, $T$ is the length of time series, and $c_i(t)$ is the number of agents who using $i$-th strategy at time $t$. 
\item[2: Detect the eigencycle]  The the eigencycles in the 2-d subspace of the game can be calculated from the eigenvector, that is, 
\begin{equation}\label{eq:sigma}
       \sigma_{mn}=\pi \cdot ||\eta_m|| \cdot ||\eta_n|| \cdot  \sin\big(\arg(\eta_m)-\arg(\eta_n)\big),
\end{equation}
Herein, $\eta_m$ is the $m$-th dimension component of the eigenvector, the 
$||\eta_m||$ is its amplitude, and $\arg(\eta_m)$ is its phase angle.
This can be verified by  time average of the angular momentum 
measurement along the time series. 
\begin{eqnarray}\label{eq:meanL}
\bar{L}_{mn}&= & \frac{1}{T}\sum_{t=0}^T x_{mn}(t) \times x_{mn}(t + 1) 
\end{eqnarray}  
Herein, $x_{mn}(t)$ is the strategy vector in the $mn$ subspace (2-d subspace), 
$T$ is the length of time series. For a 5 strategy game, the identical 2-d subspace number is 10, and the observer sample is 10 \cite{2021Qinmei}\cite{WY2020}.  The equivalent of the theoretical eigencycle  $\sigma$ in Eq. \ref{eq:sigma} and the observed angular momentum $L$ in Eq. \ref{eq:meanL} is proved referring to 
\cite{2021Qinmei,WY2020}. 
\end{description}
\subsection{Step 6: Evaluate the controller}

In the step 5 of the workflow, the controller-by-design has provided clear theoretical prediction of the controller consequence. 
So, the criteria is to test whether the predictions can be supported in significantly by data.    

We carry out the evaluation by using two approaches, simulation (see section \ref{sec:abed}) and human subject laboratory game experiment (see section \ref{sec:lab}).  
The method of agent-based simulation protocol will be shown in Appendix \ref{app:agent}. 
The method of human subject game experiment protocol  will be shown in Appendix \ref{app:humanexp}.    

Referring to the measurement of the distribution (see Eq. \ref{eq:meanrho}, denoted as $\rho_E$ for human subject game experiment and $\rho_S$ for agent-based simulation in Table \ref{tab:numRes}),  the angular momentum (see Eq. \ref{eq:meanL}, denoted as $L_E$ for human subject game experiment and $L_S$ for agent-based simulation in Table \ref{tab:numRes}) from time series,  and the theoretical prediction of Nash equilibrium (see Eq. \ref{eq:Nash_eq}, denoted as $\rho_T$  in Table \ref{tab:numRes}) and eigencycle (see Eq. \ref{eq:sigma}, denoted as $\sigma_T$  in Table \ref{tab:numRes}). The numerical results from simulation and human experiment are 
shown in Table \ref{tab:numRes}.

\begin{table}[h!]
\caption{The distribution and eigencycles of theory, simulation and human experiment}\label{tab:numResult}
\begin{center}
\begin{tabular}{rrr|rrr|rrr}
     	 \hline
\text{Distribution}~~~~~~~      	 $\rho_T^-$ & $\rho_T^o$ & $\rho_T^+$ & $\rho_E^-$ & $\rho_E^o$ & $\rho_E^+$ & $\rho_S^-$ & $\rho_S^o$ & $\rho_S^+$ \\
     	 \hline
     	 0.149 & 0.149 & 0.149 & 0.151 & 0.166 & 0.168 & 0.161 & 0.163 & 0.167 \\
     	 0.214 & 0.214 & 0.214 & 0.237 & 0.218 & 0.216 & 0.215 & 0.218 & 0.222 \\
     	 0.215 & 0.215 & 0.215 & 0.186 & 0.182 & 0.194 & 0.195 & 0.193 & 0.190 \\
     	 0.096 & 0.096 & 0.096 & 0.132 & 0.146 & 0.135 & 0.114 & 0.123 & 0.135 \\
     	 0.327 & 0.327 & 0.327 & 0.294 & 0.288 & 0.286 & 0.314 & 0.302 & 0.286 \\
     	 \hline
\text{Eigencycle }~~~~~~~~~     $\sigma^-$ & $\sigma^o$ & $\sigma^+$ & $L_E^-$ & $L_E^o$ & $L_E^+$ & $L_S^-$ & $L_S^o$ & $L_S^+$ \\
     	 \hline
     	 0.469 & 0.479 & 0.240 & 0.102 & 0.474 & 0.609  &	 0.007 & 0.012 & 0.017 \\
     	 -0.087 & -0.392 & -0.316 & -0.126 & -0.433 & -0.263  &	 -0.007 & -0.011 & -0.017 \\
     	 -0.487 & -0.258 & 0.032 & -0.178 & -0.365 & -0.306  &	 -0.015 & -0.010 & -0.013 \\
     	 0.105 & 0.171 & 0.045 & 0.202 & 0.324 & -0.040 &	 0.014 & 0.009 & 0.014 \\
     	 0.457 & 0.477 & 0.208 & 0.609 & 0.519 & 0.235 &	 0.016 & 0.016 & 0.009 \\
     	 0.224 & -0.159 & -0.358 & 0.075 & 0.053 & -0.069  &	 0.013 & -0.004 & -0.019 \\
     	 -0.212 & 0.160 & 0.389 & -0.582 & -0.098 & 0.443 &	 -0.021 & -0 & 0.027 \\
     	 0.432 & 0.387 & 0.445 & 0.332 & 0.061 & 0.303  &	 0.009 & 0.012 & 0.023 \\
     	 -0.063 & -0.302 & -0.552 & 0.151 & 0.026 & -0.332 &	 -0 & -0.006 & -0.031 \\
     	 0.169 & -0.030 & 0.118 & 0.229 & -0.252 & -0.072  &	 0.008 & -0.003 & -0.009 \\
     	 \hline

\end{tabular}
\end{center}
\label{tab:numRes}
\end{table}

\begin{table}[h!]
\caption{The correlation of eigencycles of theory ($T$), human experiment ($E$) and simulation ($S$)}
\begin{center}
\begin{tabular}{c|rrrrrrr}
     	 \hline  
corr&	& $\sigma^-$ &   $\sigma^+$ & $L_E^-$ &   $L_E^+$ & $L_S^-$  & ~~$L_S^+$ \\
(obs=10)&	 &	&	&	&	&	&	\\
     	 \hline 
$\sigma^-$ &	 &	1&	&	&	&	&	\\
$\sigma^+$ &	 &	0.300&	1&	&	&	&	\\
$L_E^-$ &	 &	0.738&	0.012&	1&	&	&	\\
$L_E^+$ &	 &	0.577&	0.773&	0.015&	1&	&	\\
$L_S^-$ &	 &	0.829&	-0.028&	0.887&	0.122&	1&	\\
$L_S^+$ &	 &	0.376&	0.910&	-0.031&	0.859&	0.039&	1\\
     	 \hline 
\end{tabular}
\end{center}
\label{tab:corr}
\end{table}

\begin{table}[h!]
\caption{The $p$-value of the linear regression over the  eigencycles of theory ($T$), human experiment ($E$) and simulation ($S$)}
\begin{center}
\begin{tabular}{c|rrrrrrr}
     	 \hline 
$p$~~~ &	& $\sigma^-$ &   $\sigma^+$ & $L_E^-$ &   $L_E^+$ & ~~$S^-$  & ~~$S^+$ \\
(obs=10)&	 &	&	&	&	&	&	\\
     	 \hline 
$\sigma^-$ &	 &	0  &	 &	 &	&	&	\\
$\sigma^+$ &	 &	  &0	 &	 &	&	&	\\
$L_E^-$ &	 &	0.015 &	 &	0 &	&	&	\\
$L_E^+$ &	 &	 &	0.009&	 &	0 &	&	\\
$L_S^-$ &	 &	0.003& &	0.001&	 &	0 &	\\
$L_S^+$ &	 &	 &	0.000&	 &	 0.001&	 &	0 \\
     	 \hline 
\end{tabular}
\end{center}
\label{tab:pvalue}
\end{table}

\subsubsection{Evaluate by agent-based evolutionary dynamics simulation}\label{sec:abed} 

Referring to the numerical results shown in Table \ref{tab:numRes} as well as the statistic results shown in Table \ref{tab:corr} and Table \ref{tab:pvalue}, we can reach the following results: 
\begin{enumerate}
\item The distribution of $[+/-]$ is identical to [o] treatment.
\item the theoretical eigencycle of $[+/-]$  can interpret the agent-based evolutionary dynamics simulation's observations  significantly ($p<0.05$). In details, 
\begin{itemize}
\item $\sigma^-$ can predict the $L_S^-$   significantly by the linear regression ($p$=0.003, $N$=10 from a 20000 rounds simulations)
\item  $\sigma^+$ can predict the $L_S^+$  significantly by the linear regression ($p$=0.000, $N$=10 from a 20000 rounds simulations)
\end{itemize} 
This means that, 
both $[-/+]$ controller can reach the desired goal.   
\item As expected by the desired goal of the controller, see the explanation of the parameters chose in Eq \ref{eq:design_goalm} and Eq. \ref{eq:design_goalp}, the $L_S^-$ vs $L_S^+$ observations have low correlation coefficient (0.0393), as shown 
in Table \ref{tab:corr}; And $L_S^-$ vs $L_S^+$ has no significant ($p>0.05$) by the linear regression, as shown 
in Table \ref{tab:pvalue} . 
This means that, as goal of the parameter chose has reached.    
\end{enumerate}


\subsubsection{Evaluate by human subject game experiment}\label{sec:lab}
 
Referring to the numerical results shown in Table \ref{tab:numRes} as well as the statistic results shown in Table \ref{tab:corr} and Table \ref{tab:pvalue} again, we can reach the following results: 
\begin{enumerate}
\item The distribution of $[+/-]$ is identical to [o] treatment.
\item The theoretical eigencycle of $[+/-]$  can interpret the human subject game experimental observations in significantly ($p<0.05$), respectively. In details, 
\begin{itemize}
\item $\sigma^-$ can predict the $L_E^-$ significantly by the linear regression ($p$=0.015, $N$=10 from 6 sessions human experiments and each session has more than  600 rounds repeated)
\item  $\sigma^+$ can predict the $L_E^+$ significantly by the linear regression ($p$=0.009, $N$=10 from 6 sessions human experiments and each session has more than  600 rounds repeated)
\end{itemize} 
This means that, 
both $[-/+]$ controllers can reach the desired goals.   
\item As expected by the desired goal of the controller, see the explanation of the parameters chose in Eq \ref{eq:design_goalm} and Eq. \ref{eq:design_goalp}, the $L_E^-$ vs $L_E^+$ observations has low correlation coefficient (0.0151), as shown 
in Table \ref{tab:corr}; And $L_E^-$ vs $L_E^+$ has no significant ($p>0.05$) by the linear regression, as shown 
in Table \ref{tab:pvalue} . 
This means that, as goal of the parameter chose has reached.  
\end{enumerate}

\subsubsection{Summary of the evaluation}\label{sec:summ}
In both of the agent-based simulation and human subjects experiment, we have following main results: \\
The theoretical eigencycle of $[+/-]$ which comes from the controlled-by-design analysis of   the $J^c$, can predicted significantly $L$ which comes from  the experimental observations as well as the agent-based simulation,  
respectively.   
 
In words, this means that, along the workflow, with the Y5 game as an example,
we have successful illustrated that, the criterion has reached ---
The dynamics structures of the two controlled treatments differ, but both can meet the their theoretical predictions respectively; And at the same time, the dynamics structure of the two controlled treatments are irrelevant although they  rooted in the same original Y5 game. In sum, we can say, to control the game dynamical structure to desired goal is reachable. 

\section{Discussion}
This paper proposes a workflow to control dynamical structure. 
By a simple and clear example game, and then by step by step, 
we illustrate that,  
the dynamical structure can be control-by-design to a desired goal can be realisable.
To the best of our knowledge, this is the first report to 
illustrate a mechanism design realisation of game dynamics structure controlling. 

\subsection{Comparison with mechanism design in game static theory}
Comparing the control-by-design in game dynamics 
with existed mechanism design in game theory concept is helpful.  
Mechanism design is a field in game theory that takes an \textbf{objectives-first approach} to designing economic mechanisms or incentives, toward desired objectives, in strategic settings, where \textbf{players act rationally}. Because it starts at the end of the game, then goes backwards, it is also called \textbf{reverse game theory}. It has broad applications, from economics and politics in such fields as market design, auction theory and social choice theory to networked-systems. Past decades has seen that, the  research and the application of mechanism design theory is extensive. 

Our game dynamics structure controller design is also an \textbf{objectives-first approach}. At the same time, the controller design itself appears as reverse optimisation too. Different from existed mechanism design, our controller design do not require the  players act rationally, but  follow the natural of human being and biology, conditional responses, myopic adaptive learning with bounded rationality  \cite{pangallo2019best}. Although the learning model appearing various, it has been notice that, the long-running social collective behaviours are \textbf{approximately outline by game replicator dynamics equation} \cite{andrade2021learning}. 

In mathematics narrative, we explain the identification of the theory and experiments of the dynamics structure in appendix (see section \ref{app:mathnarrative}) 

\subsection{Further directions of the game dynamics structure controller}
According to modern control theory, in  a linear time invariant  dynamic system,  if a given objective can be expressed as
 $d$ (eigenvalue or eigenvector), we
 can solve the unknown variable $B$ and $K$ in $J^c$.  
This is the inverse problem of the eigensystem.

In this way, when our desired goal can be expressed in $d$,
We can control the system by designing a state feedback controller, which is characterised by $B$ and $K$.  
Then theoretical results of the controlled system will show the characteristics of the desired target.


As the eigensystem bases on the Jacobian, so, when any part of eigensystem changing could lead to the changing of Jacobian, then other parts of the eigensystem changing simultaneously. As a result, when applying the control-by-design approach, carefully evaluate the consequence of control is necessary.  

Further, when a system has multi-equilibrium, controller-by-design 
of closed-loop feedback could be applied to influence the equilibrium selection. 
Such controller could provide a new way of mechanism design in game theory. 

Furthermore, when a system have multi-eigenmode,  controller-by-design 
of closed-loop feedback could be applied to influence the dynamics mode  selection. Selection of dynamics structure is not trivial. In macroeconomics, 
the dynamics structure deals with the empirical evidence which supports temporary or long-lasting business cycles. In business theory, it relates to


\bibliographystyle{plain}
\bibliography{sample}

\begin{thebibliography}{10}

\bibitem{andrade2021learning}
Gabriel~P Andrade, Rafael Frongillo, and Georgios Piliouras.
\newblock Learning in matrix games can be arbitrarily complex.
\newblock In {\em Conference on Learning Theory}, pages 159--185. PMLR, 2021.

\bibitem{2021continuous}
Volker Benndorf, Ismael Mart{\'\i}nez-Mart{\'\i}nez, and Hans-Theo Normann.
\newblock Games with coupled populations: An experiment in continuous time.
\newblock {\em Journal of Economic Theory}, 195:105281, 2021.

\bibitem{Behavioral2003}
C.F. Camerer.
\newblock {\em Behavioral game theory: Experiments in strategic interaction}.
\newblock Princeton University Press, 2003.

\bibitem{dan2014}
Timothy~N Cason, Friedman Daniel, and E.~D. Hopkins.
\newblock Cycles and instability in a rock–paper–scissors population game:
  A continuous time experiment.
\newblock {\em Review of Economic Studies}, 1:1, 2014.

\bibitem{2015feedback}
Gene~F Franklin, J~David Powell, and Abbas Emami-Naeini.
\newblock {\em Feedback control of dynamic systems}.
\newblock Pearson London, 2015.

\bibitem{fre1994}
Mark~Iosifovich Fre{\u\i}dlin and Alexander~D Wentzell.
\newblock {\em Random Perturbations of Dynamical Systems}, volume 260.
\newblock Springer-Verlag Berlin, 2012.

\bibitem{dan2016}
Daniel Friedman and Barry Sinervo.
\newblock {\em Evolutionary games in natural, social, and virtual worlds}.
\newblock Oxford University Press, 2016.

\bibitem{2015nowak}
Moshe Hoffman, Sigrid Suetens, Uri Gneezy, and Martin~A Nowak.
\newblock An experimental investigation of evolutionary dynamics in the
  rock-paper-scissors game.
\newblock {\em Scientific reports}, 5(1):1--7, 2015.

\bibitem{2019abed}
Luis~R Izquierdo, Segismundo~S Izquierdo, and William~H Sandholm.
\newblock An introduction to abed: Agent-based simulation of evolutionary game
  dynamics.
\newblock {\em Games and Economic Behavior}, 118:434--462, 2019.

\bibitem{Binmore2001Minimax}
Joe~Swierzbinski Ken~Binmore and Chris Proulx.
\newblock Does minimax work? an experimental study.
\newblock {\em The economic journal}, 2001.

\bibitem{Newton2018}
Jonathan Newton.
\newblock Evolutionary game theory: A renaissance.
\newblock {\em Games}, 9(2), 2018.

\bibitem{2010moderncontrol}
Katsuhiko Ogata.
\newblock {\em Modern control engineering}, volume~5.
\newblock Prentice hall Upper Saddle River, NJ, 2010.

\bibitem{Yoshitaka2013Minimax}
Yoshitaka Okano.
\newblock Minimax play by team.
\newblock {\em Games \& Economic Behavior}, 2013.

\bibitem{ONeill1987}
B.~O'Neill.
\newblock Nonmetric test of the minimax theory of two-person zerosum games.
\newblock {\em Proceedings of the National Academy of Sciences}, 1987.

\bibitem{1987Oneill}
B.~O'Neill.
\newblock Nonmetric test of the minimax theory of two-person zerosum games.
\newblock {\em Proceedings of the National Academy of Sciences}, 1987.

\bibitem{pangallo2019best}
Marco Pangallo, Torsten Heinrich, and J~Doyne~Farmer.
\newblock Best reply structure and equilibrium convergence in generic games.
\newblock {\em Science advances}, 5(2):eaat1328, 2019.

\bibitem{samuelson2016}
Larry Samuelson.
\newblock Game theory in economics and beyond.
\newblock {\em Journal of Economic Perspectives}, 30(4):107--30, 2016.

\bibitem{2011Sandholm}
William~H Sandholm.
\newblock {\em Population Games and Evolutionary Dynamics}.
\newblock MIT Press,, 2010.

\bibitem{2008selten}
Reinhard Selten and Thorsten Chmura.
\newblock Stationary concepts for experimental 2x2-games.
\newblock {\em American Economic Review}, 98(3):938--66, 2008.

\bibitem{smith1982evolution}
John~Maynard Smith.
\newblock {\em Evolution and the Theory of Games}.
\newblock Cambridge university press, 1982.

\bibitem{taylor1978evolutionary}
Peter~D Taylor and Leo~B Jonker.
\newblock Evolutionary stable strategies and game dynamics.
\newblock {\em Mathematical biosciences}, 40(1-2):145--156, 1978.

\bibitem{2015evolutionary}
Danielle~FP Toupo, Steven~H Strogatz, Jonathan~D Cohen, and David~G Rand.
\newblock Evolutionary game dynamics of controlled and automatic
  decision-making.
\newblock {\em Chaos: An Interdisciplinary Journal of Nonlinear Science},
  25(7):073120, 2015.

\bibitem{2020FuFeng}
Xin Wang, Zhiming Zheng, and Feng Fu.
\newblock Steering eco-evolutionary game dynamics with manifold control.
\newblock {\em Proceedings of the Royal Society A}, 476(2233):20190643, 2020.

\bibitem{wang2017}
Yijia Wang, Xiaojie Chen, and Zhijian Wang.
\newblock Testability of evolutionary game dynamics based on experimental
  economics data.
\newblock {\em Physica A: Statistical Mechanics and its Applications}, 486:455
  -- 464, 2017.

\bibitem{2012arXiv1203.2591W}
Zhijian {Wang} and Bin {Xu}.
\newblock {Evolutionary Rotation in Switching Incentive Zero-Sum Games}.
\newblock {\em arXiv e-prints}, page arXiv:1203.2591, March 2012.

\bibitem{wang2014social}
Zhijian Wang, Bin Xu, and Hai-Jun Zhou.
\newblock Social cycling and conditional responses in the rock-paper-scissors
  game.
\newblock {\em Scientific reports}, 4(1):1--7, 2014.

\bibitem{WY2020}
Zhijian Wang and Qingmei Yao.
\newblock Human social cycling spectrum.
\newblock {\em arXiv preprint arXiv:2012.03315}, 2020.

\bibitem{wiki:Dynamical_systems_theory}
Wikipedia.
\newblock {Dynamical\_systems\_theory} --- {W}ikipedia{,} the free
  encyclopedia.
\newblock
  \url{http://en.wikipedia.org/w/index.php?title=Dynamical_systems_theory},
  2021.

\bibitem{wiki:Mechanism_design}
Wikipedia.
\newblock {Mechanism\_design} --- {W}ikipedia{,} the free encyclopedia.
\newblock \url{https://en.wikipedia.org/wiki/Mechanism_design}, 2021.

\bibitem{wang2014}
Bin Xu, Shuang Wang, and Zhijian Wang.
\newblock Periodic frequencies of the cycles in $2\times2$ games: evidence from
  experimental economics.
\newblock {\em European Physical Journal B}, 87(2):46, 2014.

\bibitem{2011coyness}
Bin Xu and Zhijian Wang.
\newblock Evolutionary dynamical pattern of'coyness and philandering': Evidence
  from experimental economics.
\newblock {\em UNIFYING THEMES IN COMPLEX SYSTEMS}, 8, 2011.

\bibitem{2013cycle}
Bin Xu, Hai-Jun Zhou, and Zhijian Wang.
\newblock Cycle frequency in standard rock--paper--scissors games: evidence
  from experimental economics.
\newblock {\em Physica A: Statistical Mechanics and its Applications},
  392(20):4997--5005, 2013.

\bibitem{2021Qinmei}
Qinmei Yao.
\newblock Theoretical analysis and experiment of dynamic structure of high
  dimensional game.
\newblock
  \url{https://cdmd.cnki.com.cn/Article/CDMD-10335-1021626407.htm},doi=10.27461/d.cnki.gzjdx.2021.000847,
  2021.

\bibitem{2021Shujie}
Shujie Zhou.
\newblock Theory and experiment of dynamic structure in four strategy game.
\newblock \url{https://cdmd.cnki.com.cn/Article/CDMD-10335-1021626406.htm},
  doi=10.27461/d.cnki.gzjdx.2021.000844, 2021.

\end{thebibliography}
\section{Appendix}

\subsection{Human subject game experiment protocol\label{app:humanexp}}

The experiment was approved by the Experimental Social Science Laboratory of Zhejiang University. The data of controlled treatment ($[-/+]$-Treatment) is from the experiment carried out in Feb 2022. The data of original treatment ($o$-Treatment) is from the experiment carried out in Dec 2020 \cite{2021Qinmei}.  The authors confirms that this experiment was performed in accordance with the approved social experiments guidelines and regulations. A total number of 60 undergraduate and graduate students of Zhejiang University volunteered to serve as the human subjects of this experiment. These students were openly recruited through a web registration system. Female students were slightly more enthusiastic than male students in registering as candidate human subjects of our experiment. Since we sampled students uniformly at random from the candidate list, more female students were recruited than male students. Informed consent was obtained from all the participating human subjects.

The 60 human subjects (call also as players) were distributed into 10 populations of equal size $N$ = 6. The six players of each population carried one experimental sessions (see session organisation Table). During the game process the players sited separately in a classroom, each of which facing a computer screen. They were not allowed to communicate with each other during the whole experimental session. Written instructions were handed out to each player and the rules of the experiment were also orally explained by an experimental instructor. The rules of the experimental session are as follows:
\begin{enumerate}
\item 
Each player plays the 5-strategy game repeatedly with the same other five players.

\item 
Each player earns virtual points during the experimental session according to the payoff matrix shown in the written instruction. The rule of the tax and the reward are not known by the subjects. 

\item 
In each game round, each player competes with all the other five player as opponent.

\item 
Each player has  to make a choice among the five candidate actions “x1”, “x2”, “x3” , “x4”, “x5” . If this time runs out, the player has to make a choice immediately. After a choice has been made it can not be changed.
\end{enumerate}
 
\begin{figure}
\centering
\includegraphics[width=0.8\textwidth]{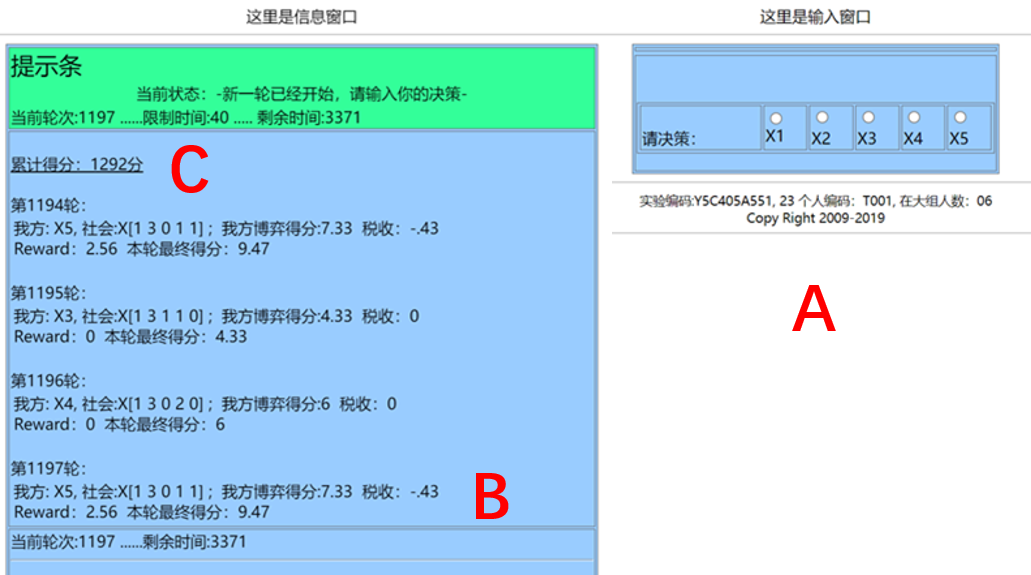}
\caption{\label{fig:expInterface} The user interface in human subject game experiment.}
\end{figure}

\begin{table}[h!]
\caption{The experiment arrange}
\begin{center}
\begin{tabular}{crrrcrr}
 \hline 
 sessionID&	Period&	subjects&~~~~~~~	&	sessionID&	Period&	subjects\\
 \hline 
15Y83n1&	591&	6&	&	15Y83p1&	629&	6\\
15Y85n2&	619&	6&	&	15Y85p2&	633&	6\\
16Y85n2&	605&	6&	&	26Y85p2&	604&	6\\
16Y85n4&	601&	6&	&	26Y85p4&	601&	6\\
19Y83n1&	603&	6&	&	19Y83p1&	669&	6\\
19Y85n2&	614&	6&	&	19Y85p2&	602&	6\\ 
 \hline
\end{tabular}
\end{center}
\label{tab:ezp}
\end{table}

During the experimental session, the computer screen of each player will show an information window and a decision window. The window on the left of the computer screen is the information window. The upper panel of this information window shows the current game round, the time limit (2 seconds in controlled  ($[-/+]$-Treatment)) of making a choice and the time left to make a choice. The color of this upper panel turns to green at the start of each game round.  After all the players  have made their decisions, the lower panel of the information window will show the player's own choice, the number of subject in each strategy (in []). 
Meanwhile, game earning, the tax, the reward and the player's own final payoff in this game round are shown in the screen. The player's own accumulated payoff is also shown. The players are asked to record their choices of each round on the record sheet in some round for checking. Each session last 20 minutes, then we have more than 600 periods records from a session. For each $[-/+]$ treatment, we have six sessions repeated. So, we have totally 3600 records in the time series.   

The window on the right of the computer screen is the decision window. It is activated only after all the players of the group have made their choices. The upper panel of this decision window lists the current game round, while the lower panel lists the five candidate actions “x1”, “x2”, “x3” , “x4”, “x5” horizontally from left to right. The player can make a choice by clicking on the corresponding action names. If a choice has not been made by a player in a period, the decision will preserve as the last decision in previous period until the player make next decision.

The reward for each player is determined by the rank, which is determined by the total number of their earning points in experiment sessions participated.  Form the highest to the lowest, each player is payed as 330, 270, 210, 150, 90 and 30 yuan RMB in controlled treatments.

\subsection{Agent-based evolutionary dynamics simulation protocol\label{app:agent}}
 Method of computer simulation to evaluate the designed controller is introduce as following. 
\begin{description}
  \item[1. Select simulation platform:] We use abed simulator \cite{2019abed}, which is widely used in the field to study evolutionary game dynamics. 
  The platform has integrated various learning rules and matching rules, and has covering mainstream dynamics model of evolutionary dynamics, which is  an ideal platform to simulate the dynamics process. The platform is of the long-running repeated game setting.  
    \item[2. Setting parameters] The simulation is under imitative protocols, in which  candidates are agents; meanwhile, the decision method is of pairwise-comparison of the strategy payoff. The complete-matching is set. These setting are follow the user guide of the platform, which system will performs like replicator dynamics shown in Eq. (\ref{eq:rp}) in large population (3000 agents) and low reversion probability (1\%) limit.  
  \item[3. Add control modular to the platform:] 
   add a modular to control the agents payoff referring to the algorithm shown in Eq. (\ref{eq:BKT}). 
  \item[4. Specify controller parameter:] 
   the modular is a state-depend feedback control 'device'. The device is specified according to the parameter vector $B$ and $K$. Various parameter assign can archive various goal. In our study case, as mentioned above, 
   \begin{center} 
   For $[-]$ controlled treatment, $B=[0,~0,~0,~0,~1]$ and $K=[1.729,	-1.611,	0.454,	1.250,	-0.400]$
  \\For $[+]$ controlled treatment, $B=[0,~0,~0,~0,~1]$ and 
  $K=[ -0.567, 1.362,  -0.963, -1.357, 0.400]$
   \end{center}
  \item[5. Conduct the simulation:]  
      In our study case, time cost for each simulation of a given parameter set is about 30 minutes in a desktop personal computer, which CPU is 8  
      GHz and the memory is 16 GB. 
  \item[6. Analysis the time series:]  Main outcome of the simulator is the time series. The time series, including the strategies density and their payoffs,  can be outputted from the platform in detail. These can be used to evaluate the performance of the controller-by-design, for example fluctuation, as well as the efficiency, profits or social welfare evolution along time. 
\end{description}

\subsection{Pole assignment for linear system}
Dynamics structure of a game is the consequence of the dynamics response of the strategy interaction of multi-agents. 
Poles assignment techniques that are used to modify the dynamic response of linear systems is one of the most studied problems in modern control theory. So, in this study, the poles assignment techniques is applied to control the dynamics structure of a game. 

Pole assignment for linear system is a basic approach in the design of control.
We apply the pole assignment of single-input linear systems using state feedback. 
We confine 
ourselves to time-invariant systems of the form
\begin{eqnarray}\label{eq:ABu}
    \dot{x}(t) &=& J x(t) + Bu(t) \\
x(0) &=& x_0
\end{eqnarray}  
and we assume that the state $x(t)$ is available. A natural control law for Eq. (\ref{eq:ABu}) is to use state feedback 
\begin{equation}\label{eq:uKx}
u =-Kx(t)  
\end{equation} 
The closed-loop system under Eq. (\ref{eq:uKx}) is then
\begin{equation}\label{eq:ABK}
    \dot{x} = (J - BK)x(t)
\end{equation} 
The closed-loop dynamics is completely determined by $(J - BK)$, and the stability of the closed-loop system as well as the rate of regulation of $x$ to zero is determined 
by the eigenvalues of $(J - BK)$, which called as  the poles of the closed-loop system.

In particular, the system Eq. (\ref{eq:ABK}) is (asymptotically) stable
if and only if all eigenvalues s of $(J - BK)$ lie in $\Re (s) < 0$. So, as far as our ability to regulate Eq. (\ref{eq:ABu}) using  (\ref{eq:uKx}) is concerned, we need to know how much control we would have over the eigenvalues of $(J - BK)$ for a given $(J, B)$ pair. The problem of finding $K$ to achieve a prescribed set of eigenvalues for $(J - BK)$ is called the pole assignment problem.

When $B=[0,0,...0,1,0,,,,0]^T$,  the control system is called as a pole assignment for single-input systems. In theory, $B$ can be arbitrary one column vector for controller design. But in our study, we need only to show the controllability, so we choose the simplest case | the single-input control, and $B=[0,0,0,0,1]$ is an example.


\subsection{Identify in mathematics narrative}\label{app:mathnarrative}
To distinguish the control-by-design approaches of game dynamics theory from existed mechanism design in game static theory, in mathematics narrative, 
we can simply summarise as follow, 
\begin{enumerate}
\item Fixed point (Nash equilibrium, or solutions concept, or stationary state) is the most concerned in existed control-by-design (or mechanism design) in game theory, 
by asking whether the equilibrium is unique,    
or which equilibrium will be selected, 
or whether the equilibrium deviate.
\item Eigenvalue of the dynamics system also concerned 
 in game theory.  Eigenvalue set is a outcome of the Jacobian at Equilibrium. 
Among the set, one of the eigenvalue should be real and reflect the value of the equilibrium value of game, meanwhile its associated vector reflect the equilibrium distribution \cite{2011Sandholm}. For a complex eigenvalue, 
the physical meaning of the real and image part differs.  
        \begin{enumerate}
        \item On real part of eigenvalue, by asking whether 
        the real part of the eigenvalue be positive 
        or negative which relating to the stability 
        of equilibrium. 
        The observation is the amplitude of the evolution trajectory 
        around the equilibrium. 
        From this point of view, 
        the dynamics theory expectations are clearly supported \cite{dan2014,wang2014social,2015nowak}. 
        \item On imaginary part  of eigenvalue, by asking 
        whether a game matrix exists cycling, 
        or which game matrix setting have higher frequency of cycle. 
        From this point of view, 
        \cite{2013cycle}\cite{dan2014} has noticed potential impact of
        the imaginary part and reported the existence of cycle, 
        with the net angle or net cycle counting. Meanwhile,
        \cite{wang2014} has provided the periodic frequency cycling evidence in 12 treatments of $2\times 2$ games experiment reported in \cite{2008selten};
         \cite{2012arXiv1203.2591W}
        provides the angular momentum evidence in experiments data in \cite{Binmore2001Minimax}. All these support
        the dynamics theory of the replicator dynamics class. 
        \end{enumerate} 
\item Complex eigenvector of the  dynamics  system is also concerned. 
We have seen that, the eigenvector structure determines 
the dynamical structure in high dimensional games:
\begin{enumerate}
\item \cite{WY2020} provides evidence from O'Neill game experiment conducted 
by O'Neill in 1987 \cite{1987Oneill}, by Binmore et al 2001 \cite{Binmore2001Minimax} and by Okano in 2013  \cite{Yoshitaka2013Minimax} respectively. In this game, the eight components of 
eigenvectors can predict 28 2-d subspace cycling behaviours. In \cite{WY2020},
the authors develop an observation, namely eigencycle set,  
which is determined by the eigenvector deduced from dynamics theory; And at the same time, the eigencycle set are mathematically equivalent to angular momentum measurement \cite{wang2017} in time series .
\item \cite{2021Qinmei} provides eigencycle evidence of 
a $5\times 5$ game with unique mixed strategy Nash equilibrium, 
in which there exists only one pair of complex eigenvector
with five components which determines the cycles the 10 2-d sub-spaces. 
\item \cite{2021Shujie} provides evidences
from two treatments of $4\times 4$ game
with various game matrix, 
where the cyclic manifold from dynamics theory is confirmed.
\end{enumerate}
These findings show that, the logic chain --- 
dynamics equations, its equilibrium, Jacobian, 
eigenvalue, eigenvector  and their components --- 
can provides systematically and accurately 
insight the real game dynamics behaviour.  
Here, from the narrative of paradigm 
shift from game statics theory, 
the eigenvector plays a cruel role. 

Then changing the eigenvector is not trivial.
For more detail, changing the eigenvector structure has two approaches,
\begin{enumerate}
\item The first is by control the incentive of game, or saying to change the payoff matrix of game, 
to detect whether the dynamics structure change. 
Form this point of view, the experiments in \cite{2021Shujie} 
have illustrated an example by controlling one element 
in a  $4 \times 4$ (16 elements) payoff matrix game 
to change the eigenvector and then the dynamics structure
identified by the eigencycle measurement, 
in which  the game dynamics theory 
was supported significantly.
\item The second is using feedback control approach, which 
borrowing from the modern control theory. 
Having above outline from fixed point 
to eigenvalue to eigenvector to eigencycle,
it is not surprise  
to seek the dynamics structure changed by controller-by-design to desired eigenvector.  
\end{enumerate} 

\end{enumerate}
 
\subsection{Controller parameters selection}\label{app:designgoal}

This subsection is an explanation for the section \ref{sec:assigngoal} in main text. 

The aim of this research is to illustrate the approach --- pole assignment for linear system with single-input --- 
can control game dynamics structure. 

On control objective, 
as the dynamics structure mainly depends on the complex eigenvalues and the associated eigenvectors, and as pole assignment for linear system can control the eigenvalues, we can choose to control the real part of the complex eigenvalue $\lambda^o$. 

On optimal parameters selection, we refer to: 
\begin{itemize}
\item Change one parameter as small as possible, so (1) we choose one channel (which belongs single-input control approach) control approach; (2) we choose only the real part of the complex eigenvalue $\lambda^o(3)$ in Eq. \ref{eq:lambda_o} to  be modified by $b$. The technology to modify $b$ belonging to pole assignment control approach in modern control theory;
\item Observe significantly different in experiment, that is make sure that the two control goals are reached. The two goals are: 
\begin{enumerate}
\item the theoretical eigencycle $\sigma^-$ can interpret $L$ observed in $[-]$ treatment statistical significant;
And at the same time, $\sigma^-$ can not interpret $L$ observed in $[+]$ treatment. 
\item $\sigma^+$ can interpret $L$ observed in $[+]$ treatment in statistical significant; And at the same time, can not interpret $L$ observed in $[-]$ treatment. 
\end{enumerate} 
\end{itemize}

\begin{figure}[h!]
\centering
\includegraphics[width=1\textwidth]{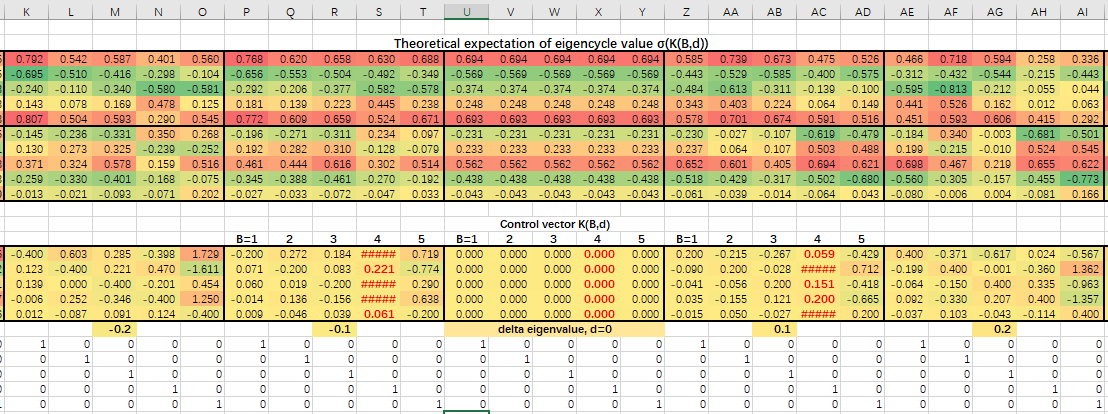} 
\caption{Results of the $K$ and eigencycle value referring to various $b$ and $B$. From left to right, the 25 columns are organised as $\big(b=-0.2, ~~B=[1,0,0,0,0]\big)^T$, $\big(b=-0.2, ~~B=[0,1,0,0,0]\big)^T$, $\big(b=-0.2, ~~B=[0,0,1,0,0]\big)^T$, ... $\big(b=-0.1, B=[1,0,0,0,0]\big)^T$, $\big(b=-0.1, ~~B=[0,1,0,0,0]\big)^T$, ..., $\big(b=+0.2, ~~B=[0,0,0,0,1]\big)^T$. The coloured rows in middle block, which having 5 rows, provides the control parameter $K$ values. The bottom block, which having 5 rows,  provides the channel $B$ value which indicates the single-input vectors. The 10 rows in the top block provides the theoretical eigencycle values of the control parameters $K$ and $B$.   Notice that, the eigencycle value $\sigma^-$,  $\sigma^o$ and $\sigma^+$ shown in Table \ref{tab:numResult} is normalised by the root of the sum of the square of the components shown in the 5th,  the 15th and 25th column of the top block, respectively.  \label{fig:designgoal}} 
\end{figure}

\begin{figure}[h!]
\centering 
\includegraphics[width=1\textwidth]{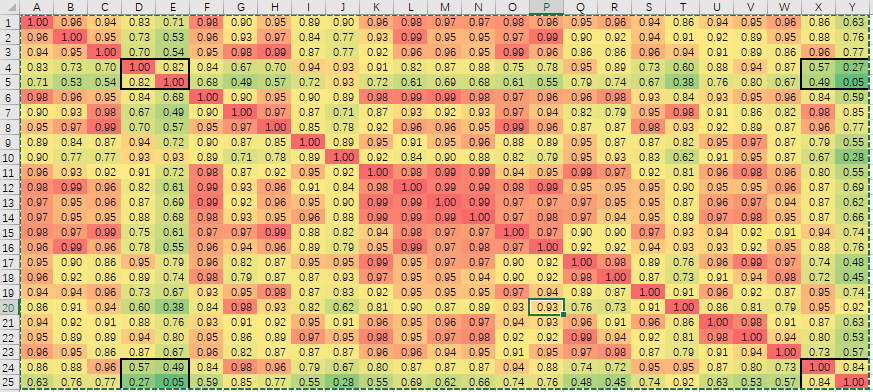}
\caption{The correlation coefficient of the 25 eigencycle sets. From top to bottom, as well as from the left to right, the 25 candidates are organised as $\big(b=-0.2, ~~B=[1,0,0,0,0]\big)^T$, $\big(b=-0.2, ~~B=[0,1,0,0,0]\big)^T$, $\big(b=-0.2, ~~B=[0,0,1,0,0]\big)^T$, ... $\big(b=-0.1, B=[1,0,0,0,0]\big)^T$, $\big(b=-0.1, ~~B=[0,1,0,0,0]\big)^T$, ..., $\big(b=+0.2, ~~B=[0,0,0,0,1]\big)^T$.
We choose the lowest value condition, where correlated value is 0.05 (25-th row, 5-th column), that is $B=[0~0~0~0~1]$ and $b=\pm 0.2$ condition. These two conditions are exactly the $[-/+]$-treatment we focus in this study.\label{fig:designgoa2}} 
\end{figure}

In details, we select the parameters by following steps
\begin{enumerate}
\item We scan parameter $b$ near 0 with step 0.1 (there are five candidates $b=\{-0.2, -0.1,0,0.1,0.2\}$);  We scan also 
 the five candidates $B$ ($B=[1,0,0,0,0]$, $B=[0,1,0,0,0]$, $B=[0,0,1,0,0]$, $B=[0,0,0,1,0]$, $B=[0,0,0,0,1]$). Then for each of the 25 candidates, we can calculates their $K$, respectively.
When calculating $K$, we use the pole assignment approach. The results are the 5 rows shown in the middle block in figure \ref{fig:designgoal}.
\item Then with the $K$,  the controlled Jacobian ${J^c}$ can be obtained with Eq.~\ref{eq:jc1}.    
And then with ${J^c}$ the related eigencycle values of 10 subspace can be obtained. 
The results are the 10 rows shown in the top block in figure \ref{fig:designgoal}.
\item Then, we can calculate the the correlation coefficient of the 25 eigencycle sets. We choose the lowest value of the correlation coefficient condition.  
In figure \ref{fig:designgoa2}, we show all the correlated coefficient,
in which the lowest is 0.05, 
that is $B=[0~0~ 0~ 0~ 1]$ and $b=\pm 0.2$ condition. 
The results are shown the black box in figure \ref{fig:designgoa2}.
\end{enumerate}
As the results, $B=[0~0~ 0~ 0~ 1]$ and $b=\pm 0.2$ are labelled as the $[+/-]$ treatments in main text Section \ref{sec:assigngoal}.

\end{document}